\documentclass[pre, twocolumn, showkeys, amsmath, amssymb]{revtex4}

\usepackage{graphicx}

\begin{document}

\title{Temporal Differentiation and the Optimization of System Output}

\author{Emmanuel Tannenbaum}
\email{etannenb@gmail.com}
\affiliation{Department of Chemistry, Ben-Gurion University of the Negev,
Be'er-Sheva 84105, Israel}

\begin{abstract}

This paper develops a set of simplified dynamical models with which to explore
the conditions under which temporal differentiation leads to optimized system
output.  By temporal differentiation, we mean a division of labor whereby different 
subtasks associated with performing a given task are done at different times.  The 
idea is that, by focusing on one particular set of subtasks at a time, it is possible
to increase the efficiency with which each subtask is performed, thereby allowing for
faster completion of the overall task.  For this paper, we consider a process whereby
some resource is converted into some final product in a series of three agent-mediated
steps.  Temporal differentiation is incorporated by allowing the agents to oscillate 
between performing the first two steps and performing the last step.  We find that
temporal differentiation is favored when the number of agents is small, and when the
process intermediates have a much longer lifetime than the original resource.  
Within the framework of biological systems, we argue that these results provide a possible 
evolutionary basis for the emergence of sleep, and also of distinct REM and non-REM sleep states.  We also
discuss our use of a three-step model.  Briefly, in order for temporal differentiation to
increase product output in a mean-field description of resource metabolism, it is necessary
for temporal differentiation to have a nonlinear effect on individual process rates.  For stochastic
models, we argue that temporal differentiation can increase product output even in 
fundamentally linear systems. 

\end{abstract}


\maketitle

\section{Introduction}

Differentiation and the division of labor is a ubiquitous phenomenon characterizing the emergence
of complex systems.  Different enzymes, nucleic acids and other biopolymers are involved in the
proper function of a living cells.  In multicellular organisms, cells differentiate and 
specialize in the performance of one or a few tasks.  At higher levels of complexity, multicellular
organisms, e.g. humans, can themselves form highly differentiated structures (a modern networked economy),
where each organism performs one or a few tasks \cite{ORGDIV1, ORGDIV2, BACTDIV, INSECTDIV1, INSECTDIV2,
ECONDIV1, ECONDIV2, ECONDIV3, ECONDIV4, ECONDIV5, ECONDIV6, ECONDIV7, ECONDIV8}.

As a result of the ubiquity of the division of labor in biology, considerable experimental and theoretical
work has been devoted to understanding both its genetic basis, and the selection pressures that give
rise to such behaviors.  In a general sense, division of labor is favored when transport costs associated 
with delivering process intermediates to the appropriate agents are small.  Therefore, division of labor is 
generally favored at high population densities, though this may not always be the case \cite{ECONDIV5, DIFFTANN}.

In this paper, we wish to discuss another form of differentiation, specifically {\it temporal differentiation}.
Temporal differentiation refers to a division of labor where a given task is broken up into several
subtasks, and the various subtasks are performed at different times.  That is, with temporal differentiation,
a given set of agents performs all the subtasks associated with a given task.  However, these agents concentrate
their efforts on one set of subtasks for a certain period of time, and then concentrate their efforts on
another set of subtasks for another period of time.  This is in contrast to the ``standard'' picture of division
of labor, whereby all subtasks associated with a given task are performed simultaneously by different sets of
agents.

Temporal differentiation, although a perhaps less obvious form of division of labor, is nevertheless also 
a ubiquitous phenomenon.  At the level of task completion by humans, it is quite common that various tasks are 
often done in intermittent blocs.  Examples include paying of bills, housekeeping chores, and the procurement
of food.  Such forms of temporal differentiation are likely prevalent in other organisms, since such a temporally
differentiated labor strategy is a natural approach in many contexts for optimizing system efficiency.

Another, more subtle form of temporal differentiation is the phenomenon of sleep.  Sleep is prevalent
in organisms with highly complex central nervous systems.  It is characterized by periods of high levels of alertness,
followed by periods during which the brain goes into an unconscious state.  

One theory, due to Crick and Mitchison, for the existence of sleep, is that sleep is a time when the brain engages 
in various garbage collection activities \cite{CRICKHYP}.  More specifically, sleep is a time when the brain sorts through and consolidates
information accumulated from the previous period of wakefulness \cite{MEMHYP1, MEMHYP2, MEMHYP3, MEMHYP4, MEMHYP5, MEMHYP6}.  Increasing 
evidence suggests that Crick and Mitchison's basic hypothesis may be correct \cite{FOS}.  Nevertheless, while this ``garbage collection'' hypothesis 
provides an explanation for what happens during sleep, it does not explain the selection pressures driving the 
emergence of this phenomenon.

In two recent papers \cite{SLEEPHYP, SLEEPTANK}, the author suggested that sleep emerges because, in the presence of a day-night cycle, it is optimal
for a highly complex brain to engage in information collection activities during the day, when light information is
most available, and then to engage in information consolidation (i.e. ``garbage collection'') activities at night, when
light information is far less available.  The idea is that, by concentrating on information collection when it is available,
and information consolidation when external information is less available, the brain can process an optimal amount of
information in a given amount of time, which presumably confers a survival advantage to the organism.

However, such an explanation is incomplete, since some organisms are nocturnal.  Further, what is interesting about the sleep
state itslef is that it is divided into distinct, alternating cycles of REM and non-REM sleep (where REM stands for
``Rapid Eye Movement'') \cite{CRICKHYP, MEMHYP1, SLEEPDEPRIVE}.  However, not all sleeping organisms exhibit REM and non-REM sleep.  
In particular, the organism {\it Tachyglossus aculeatus}, a representative of the earliest branch of mammalian evolution (the monotremes), 
combines both REM and non-REM sleep into one sleep state \cite{OLDSLEEP}.  This suggests that REM and non-REM sleep are not fundamental to sleep itself, but rather
emerged via the differentiation of a single, older sleep state.  

Therefore, it is possible that temporal differentiation of various brain tasks leads to optimized brain function in 
higher organisms, independent of any external day-night regulation cycle.  The existence of a day-night cycle simply
regulates the optimal start times for each task period.  Furthermore, it is also possible that temporal differentiation is
favored the more work needs to be completed within a given time period.  In the context of REM and non-REM
sleep, it is possible that mammals with simpler brains can perform the information consolidation tasks in one sleep
state.  However, for larger, more complex mammalian brains, the amount of information consolidation to be performed
becomes sufficiently large that it becomes more efficient to divide the various information consolidation subtasks into
two distinct sleep states.

Motivated by these various considerations, we present a highly simplified model for the processing of some resource
into a final product.  The processing of this resource occurs in three separate steps, each mediated by an agent.
Because our model adopts the notation and structure of the theory of chemical reaction kinetics, the various agents
involved in each subtask are represented as enzyme catalysts.  

When there is no temporal differentiation, all three subtasks are performed simultaneously,
and the fraction of enzymes assigned to each task is constant in time.  With temporal differentiation in the context
of our model, the enzymes oscillate between performing the first two tasks for a certain amount of time, and then
the third task for another amount of time.  We use a three-subtask model because, in the mean-field approach that
we adopt here, a minimum of three subtasks is necessary in order to achieve the nonlinear dependence of task rate
on enzyme number required to give an advantage to a temporally differentiated labor strategy.  For stochastic models,
this fundamentally nonlinear dependence of task rate on enzyme number is not necessary, an issue which we will discuss
later.

Within the framework of our model, we find that temporal differentiation is favored when the amount of available agents
is at intermediate values compared to the rate of resource input, and when the intermediate products have a long lifetime compared
to the starting resource.  It therefore makes sense to focus entirely on one set of tasks for a time, to convert 
as much resource as possible into a long-lived intermediate, and then switch to processing the intermediate.

This paper is organized as follows:  In the following section (Section II), we develop our three-process model,
and derive the limiting form of the model when the total number of enzymes is small.  In Section III, we go on 
to solve our model with and without temporal differentiation.  In Section IV, we compare both the temporally 
differentiated and non-differentiated strategies, and determine the regimes where one strategy is expected to be favored 
over the other.  In Section V we discuss our results in the context of sleep and other examples of temporal differentiation 
in biology.  Finally, in Section VI we present our main conclusions.  We also describe some shortcomings of our model, 
and how our model may be further developed in future work.

\section{The Model}

In this section, we introduce our three process model, whereby an external resource is converted into a final
product in a series of three steps.  We set up dynamical equations governing the production of final product
for both the temporally differentiated and non-differentiated cases.

\subsection{Definition of the model}

Our model consists of some compartment of fixed volume $ V $, into which flows a resource, denoted
$ R_1 $, at some fixed rate $ f_R $.  This resource is processed into a final product, denoted $ P $,
via a series of three agent-mediated, or, in the language of chemical kinetics, enzyme-catalyzed, steps.

In the first step, an enzyme, denoted $ E_1 $, binds to $ R_1 $, and then converts $ R_1 $ into an
intermediate $ R_2 $.  In the second step, an enzyme, denoted $ E_2 $, binds to $ R_2 $, and then converts
$ R_2 $ into an intermediate $ R_3 $.  In the third and final step, an enzyme, denoted $ E_3 $, binds to $ R_3 $,
and then converts $ R_3 $ into the final product $ P $.  

Furthermore, the resource $ R_1 $ and intermediates $ R_2 $, $ R_3 $ have finite lifetimes in the compartment,
defined by first-order decay constants $ k_{D, 1} $, $ k_{D, 2} $, and $ k_{D, 3} $, respectively.  These decay
terms can be due to various factors, such as diffusion out of the compartment, or simply the physical decay
of the components themselves.  In the context of networked systems and data processing, these decay constants
can also correspond to a finite lifetime during which an information packet is relevant (say stock information 
that is used by an investor to decide whether or not to invest in a given stock).

In the language of chemical kinetics, the set of reactions in the compartment is given by,
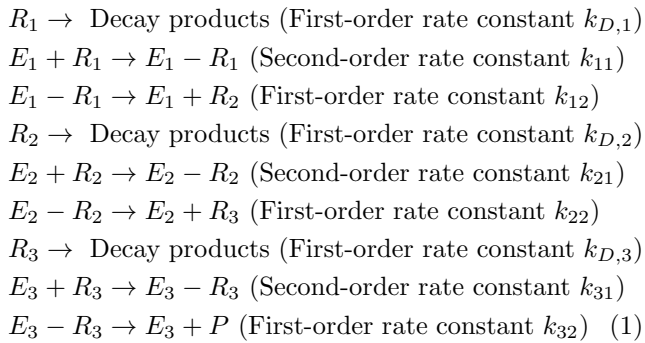
\begin{eqnarray}
&   &
R_1 \rightarrow \mbox{ Decay products} \mbox{ (First-order rate constant $ k_{D, 1} $)}
\nonumber \\
&   &
E_1 + R_1 \rightarrow E_1-R_1 \mbox{ (Second-order rate constant $ k_{11} $)}
\nonumber \\
&   &
E_1-R_1 \rightarrow E_1 + R_2 \mbox{ (First-order rate constant $ k_{12} $)}
\nonumber \\
&   &
R_2 \rightarrow \mbox{ Decay products} \mbox{ (First-order rate constant $ k_{D, 2} $)}
\nonumber \\
&   &
E_2 + R_2 \rightarrow E_2-R_2 \mbox{ (Second-order rate constant $ k_{21} $)}
\nonumber \\
&   &
E_2-R_2 \rightarrow E_2 + R_3 \mbox{ (First-order rate constant $ k_{22} $)}
\nonumber \\
&   &
R_3 \rightarrow \mbox{ Decay products} \mbox{ (First-order rate constant $ k_{D, 3} $)}
\nonumber \\
&   &
E_3 + R_3 \rightarrow E_3-R_3 \mbox{ (Second-order rate constant $ k_{31} $)}
\nonumber \\
&   &
E_3-R_3 \rightarrow E_3 + P \mbox{ (First-order rate constant $ k_{32} $)}
\end{eqnarray}

Letting $ n_{R_i} $ denote the number of particles of resource $ R_i $, 
$ n_{E_i} $ the number of particles of enzyme $ E_i $, and $ n_{E_i R_i} $
the number of particles of $ E_i-R_i $, where $ i = 1, 2, 3 $, we obtain
the following system of differential equations for the time evolution of
the numbers of the various components in the compartment:

\begin{eqnarray}
&   &
\frac{d n_{R_1}}{dt} = f_R - \frac{k_{11}}{V} n_{E_1} n_{R_1} - k_{D, 1} n_{R_1}
\nonumber \\
&   &
\frac{d n_{E_1 R_1}}{dt} = \frac{k_{11}}{V} n_{E_1} n_{R_1} - k_{12} n_{E_1 R_1}
\nonumber \\
&   &
\frac{d n_{R_2}}{dt} = k_{12} n_{E_1 R_1} - \frac{k_{21}}{V} n_{E_2} n_{R_2} - k_{D, 2} n_{R, 2}
\nonumber \\
&   &
\frac{d n_{E_2 R_2}}{dt} = \frac{k_{21}}{V} n_{E_2} n_{R_2} - k_{22} n_{E_2 R_2}
\nonumber \\
&   &
\frac{d n_{R_3}}{dt} = k_{22} n_{E_2 R_2} - \frac{k_{31}}{V} n_{E_3} n_{R_3} - k_{D, 3} n_{R, 3}
\nonumber \\
&   &
\frac{d n_{E_3 R_3}}{dt} = \frac{k_{31}}{V} n_{E_3} n_{R_3} - k_{32} n_{E_3 R_3}
\end{eqnarray}

We also define the quantities $ n_1 $, $ n_2 $, and $ n_3 $ via,
\begin{eqnarray}
&   &
n_1 = n_{E_1} + n_{E_1 R_1}
\nonumber \\
&   &
n_2 = n_{E_2} + n_{E_2 R_2}
\nonumber \\
&   &
n_3 = n_{E_3} + n_{E_3 R_3}
\end{eqnarray}
so that $ n_1 $, $ n_2 $, and $ n_3 $ denote the total amount of $ E_1 $, $ E_2 $, and $ E_3 $
respectively at some given time.

Note that our model is essentially a mean-field description of the dynamics inside the compartment,
since we are assuming that the amount of each component can take on any nonnegative real number.

\subsection{Temporal differentiation in the three-enzyme compartment model}

Temporal differentiation can occur in our model if we do not assume that $ n_1 $, $ n_2 $,
and $ n_3 $ are fixed, but rather can oscillate in time.  For the purposes of this paper,
we will assume that $ n_1 $, $ n_2 $, and $ n_3 $ oscillate in such a way that $ n_1 = n_1^{+} $,
$ n_2 = n_2^{+} $, and $ n_3 = n_3^{-} $ over some time period of length $ T_1 $, followed by
a time period of length $ T_2 $ where $ n_1 = n_1^{-} \leq n_1^{+} $, $ n_2 = n_2^{-} \leq n_2^{+} $,
and $ n_3 = n_3^{+} \geq n_3^{-} $.  We assume that the total number of enzymes remains fixed,
however, so that $ n_1^{+} + n_2^{+} + n_3^{-} = n_1^{-} + n_2^{-} + n_3^{+} $.

Essentially, if we switch from an enzyme-based viewpoint to an agent-based viewpoint, our model assumes
that agents can switch from one set of tasks to another.  In this model, the agents alternate between
focusing on the first two processes and the third process.  In the chemical kinetics notation, we
have,
\begin{equation}
E_{1/2} \leftrightarrow E_3
\end{equation}

For the purposes of this paper, we will assume that the rate constant of this ``task-switching''
reaction is $ \infty $.  Therefore, each agent can instantaneously switch from one task to another,
if it so chooses.  Toward the end of this paper we will speculate on the effect that a finite
task-switching rate has on the results presented here.

\subsection{Limiting forms of the model}

We will now study how our model behaves when $ n_1 + n_2 + n_3 $ and $ k_{D, 3} $ may each
be regarded as ``small'' in some sense.  The criterion for smallness will be defined later,
once we have established the behavior of the model in these regimes.

To begin, we note from the previous subsection that the various enzyme numbers fluctuate in time.  
More precisely, there exist $ T_1, T_2 > 0 $ such that for every integer $ s $, the total enzyme 
numbers for each enzyme are at $ n_1^{+} $, $ n_2^{+} $, $ n_3^{-} $ respectively, during the time interval
$ [s (T_1 + T_2), s (T_1 + T_2) + T_1] $, while during the time interval $ [s (T_1 + T_2) + T_1,
(s + 1) (T_1 + T_2)] $, the enzyme numbers are at $ n_1^{-} $, $ n_2^{-} $, and $ n_3^{+} $.

If we define,
\begin{equation}
\bar{n}_{1/2/3} = \frac{n_{1/2/3}^{+} + n_{1/2/3}^{-}}{2}
\end{equation}
and,
\begin{equation}
\lambda_{1/2/3} = \frac{n_{1/2/3}^{+} - n_{1/2/3}^{-}}{n_{1/2/3}^{+} + n_{1/2/3}^{-}}
\end{equation}
then it may be readily shown that,
\begin{equation}
n_{1/2/3}^{\pm} = \bar{n}_{1/2/3} (1 \pm \lambda_{1/2/3})
\end{equation}

For fixed values of $ \lambda_1 $, $ \lambda_2 $, we wish to develop a form for the first four 
equations assuming that $ \bar{n}_{1} $, $ \bar{n}_2 $ are small.  The overall strategy is
as follows:  Because $ R_3 $ is the intermediate that feeds into the third task, our goal is
to determine the rate of production of $ R_3 $ when $ \bar{n}_1 $, $ \bar{n}_2 $ are small.
By this we mean that we seek to determine, with respect to $ \bar{n}_1 $ and $ \bar{n}_2 $,
the lowest-order term contributing to the production rate of $ R_3 $.

When $ \bar{n}_1 = 0 $, we have that $ n_{E_1 R_1} = n_{E_1} = n_{R_2} = n_{E_2 R_2} = n_{E_2} = 0 $.
If we let $ n_{R_1, 0} $ denote $ n_{R_1} $ when $ \bar{n}_1 = 0 $, we obtain,
\begin{equation}
\frac{d n_{R_1, 0}}{dt} = f_R - k_{D, 1} n_{R_1, 0}
\end{equation}

Differentiating the second equation with respect to $ \bar{n}_1 $, and setting $ n_{E_1 R_1, 1} = 
(\partial n_{E_1 R_1}/\partial \bar{n}_1)_{\bar{n}_1 = 0} $, we obtain,
\begin{equation}
\frac{d n_{E_1 R_1, 1}}{d t} = 
\frac{k_{11}}{V} (1 \pm \lambda_1) n_{R_1, 0} - (\frac{k_{11}}{V} n_{R_1, 0} + k_{12}) n_{E_1 R_1, 1}
\end{equation}

Differentiating the third equation with respect to $ \bar{n}_1 $, and setting $ n_{R_2, 1} =
(\partial n_{R_2}/\partial \bar{n}_1)_{\bar{n}_1 = \bar{n}_2 = 0} $, we obtain,
\begin{equation}
\frac{d n_{n_{R_2, 1}}}{dt} = k_{12} n_{E_1 R_1, 1} - k_{D, 2} n_{R_2, 1}
\end{equation}

Note that $ n_{E_2} = n_{E_2 R_2} = 0 $ when $ \bar{n}_2 = 0 $.  Therefore, derivatives
of these quantities that only involve $ \bar{n}_1 $ will be $ 0 $ when evaluated at
$ (\bar{n}_1, \bar{n}_2) = (0, 0) $.  So, the lowest-order derivative at $ n_{E_2 R_2} $ that is possibly
non-vanishing at $ (\bar{n}_1, \bar{n}_2) = (0, 0) $ is $ \partial^2 n_{E_2 R_2}/(\partial \bar{n}_1 \partial \bar{n}_2) $.
Defining $ n_{E_2 R_2, 1} = (\partial^2 n_{E_2 R_2}/(\partial \bar{n}_1 \partial \bar{n}_2))_{(\bar{n}_1, \bar{n}_2) = (0, 0)} $,
we obtain,
\begin{equation}
\frac{d n_{E_2 R_2, 1}}{dt} = \frac{k_{21}}{V} (1 \pm \lambda_2) n_{R_2, 1} - k_{22} n_{E_2 R_2, 1}
\end{equation}

Note that the lowest order term contributing to the production rate of $ R_3 $ is given by
$ n_{E_2 R_2, 1} \bar{n}_1 \bar{n}_2 $.

For the first four equations, we obtain the linearized system,
\begin{eqnarray}
&   &
\frac{d n_{R_1, 0}}{dt} = f_R - k_{D, 1} n_{R_1, 0}
\nonumber \\
&   &
\frac{d n_{E_1 R_1, 1}}{dt} = \frac{k_{11}}{V} (1 \pm \lambda_1) n_{R_1, 0} - (\frac{k_{11}}{V} n_{R_1, 0} + k_{12}) n_{E_1 R_1, 1}
\nonumber \\
&   &
\frac{d n_{R_2, 1}}{dt} = k_{12} n_{E_1 R_1, 1} - k_{D, 2} n_{R_2, 1}
\nonumber \\ 
&   &
\frac{d n_{E_2 R_2, 1}}{dt} = \frac{k_{21}}{V} (1 \pm \lambda_2) n_{R_2, 1} - k_{22} n_{E_2 R_2, 1}
\end{eqnarray}

For the final two equations, we assume that $ k_{D, 3} = 0 $, giving,
\begin{eqnarray}
&   &
\frac{d n_{R_3}}{dt} = k_{22} n_{E_2 R_2} - \frac{k_{31}}{V} n_{E_3} n_{R_3}
\nonumber \\
&   &
\frac{d n_{E_3 R_3}}{dt} = \frac{k_{31}}{V} n_{E_3} n_{R_3} - k_{32} n_{E_3 R_3}
\end{eqnarray}

We will discuss the assumption of $ k_{D, 3} = 0 $ later in this paper.

\section{Long time behavior of the model}

In the absence of temporal differentiation, the values of $ n_1 $, $ n_2 $, $ n_3 $ remain constant, and so
we expect the dynamics to evolve to a steady-state.  With temporal differentiation, the values of $ n_1 $,
$ n_2 $, $ n_3 $ oscillate between two sets of values over a well-defined time period $ T_1 + T_2 $.  Therefore,
in this situation, we cannot expect the dynamics to settle into a steady-state solution.  However, we can
expect the dynamics to settle into a periodic solution.

In this section, we therefore consider the long-term behavior of our model with and without temporal differentiation.
Without temporal differentiation, that is, assuming that $ \lambda_1 = \lambda_2 = \lambda_3 = 0 $, we look for steady-state
solutions to the dynamical system.  With temporal differentiation, we look for periodic solutions to the dynamical
system.  We will consider the case where $ \lambda_1 = \lambda_2 = \lambda_3 = 1 $.  Here, the enzymes devote themselves entirely
to the first two tasks over a time interval of length $ T_1 $, and then devote themselves to the third task over a time
interval of length $ T_2 $.

\subsection{Case 1:  $ \lambda_1 = \lambda_2 = \lambda_3 = 0 $}

When $ \lambda_1 = \lambda_2 = \lambda_3 = 0 $, there is no fluctuation in the total enzyme numbers for either $ E_1 $, $ E_2 $,
or $ E_3 $.  Therefore, the long-time solution is simply a steady-state solution.  Setting
the left-hand sides of the four linearized equations to $ 0 $ gives, 
\begin{eqnarray}
&   &
n_{R_1, 0} = \frac{f_R}{k_{D, 1}}
\nonumber \\
&   &
n_{E_1 R_1, 1} = \frac{\frac{f_R}{k_{12}}}{\frac{f_R}{k_{12}} + \frac{k_{D, 1}}{(k_{11}/V)}}
\nonumber \\
&   &
n_{R_2, 1} = \frac{k_{12}}{k_{D, 2}} \frac{\frac{f_R}{k_{12}}}{\frac{f_R}{k_{12}} + \frac{k_{D, 1}}{(k_{11}/V)}}
\nonumber \\
&   &
n_{E_2 R_2, 1} = \frac{(k_{21}/V)}{k_{D, 2}} \frac{k_{12}}{k_{22}}
\frac{\frac{f_R}{k_{12}}}{\frac{f_R}{k_{12}} + \frac{k_{D, 1}}{(k_{11}/V)}}
\end{eqnarray}

If we define $ A $ and $ B $ via,
\begin{eqnarray}
&   &
A = \frac{(k_{11}/V)}{k_{D, 1}} f_R
\nonumber \\
&   &
B = \frac{(k_{11}/V)}{k_{D, 1}} f_R + k_{12}
\end{eqnarray}
then we obtain that the rate of production of $ R_3 $ is $ k_{12} (k_{21}/V)/k_{D, 2} (A/B) \bar{n}_1
\bar{n}_2 $.

Now, the maximal production rate of $ P $ is obtained when $ n_{R_3} = \infty $.  The reason for this is that 
the rate of the third binding step is then infinite, so that the production rate of $ P $ is only limited by the rate at which
the enzyme $ E_3 $ can convert $ R_3 $ into $ P $.  

When $ n_{R_3} = \infty $, the rate of production of $ P $ is given by $ k_{32} n_{E_3 R_3} = k_{32} \bar{n}_3 $.
At steady-state, the rate of production of $ R_3 $ must equal the rate of production of $ P $ (since the production
rate of $ P $ is equal to the consumption rate of $ R_3 $ when $ k_{D, 3} = 0 $).  Therefore, at steady-state,
\begin{equation}
k_{12} \frac{(k_{21}/V)}{k_{D, 2}} (\frac{A}{B}) \bar{n}_1 \bar{n}_2 = k_{32} \bar{n}_3
\end{equation}
Now, defining $ n = \bar{n}_1 + \bar{n}_2 + \bar{n}_3 $, $ \alpha = (\bar{n}_1 + \bar{n}_2)/n $,
$ \beta = \bar{n}_1/(\bar{n}_1 + \bar{n}_2) $, we have, at steady-state,
\begin{equation}
\frac{d n_{P}}{dt} = k_{12} \frac{(k_{21}/V)}{k_{D, 2}} (\frac{A}{B}) \alpha^2 \beta (1 - \beta) n^2 = k_{32} (1 - \alpha) n
\end{equation}

Therefore, the steady-state production rate of $ P $ is maximized when $ \beta = 1/2 $, and so we
wish to solve,
\begin{equation}
\alpha^2 + \frac{1}{\gamma n} \alpha - \frac{1}{\gamma n} = 0
\end{equation}
where,
\begin{equation}
\gamma \equiv \frac{k_{12} \frac{(k_{21}/V)}{k_{D, 2}} (\frac{A}{B})}{4 k_{32}}
\end{equation}

This gives,
\begin{equation}
\alpha = \frac{1}{2} \frac{1}{\gamma n} [-1 + \sqrt{1 + 4 \gamma n}]
\Rightarrow
1 - \alpha = \frac{1 + 2 \gamma n - \sqrt{1 + 4 \gamma n}}{2 \gamma n}
\end{equation}

The steady-state production rate of $ P $ is then given by $ k_{32} (1 - \alpha) n $.  To distinguish
this value of $ \alpha $ from the value of $ \alpha $ we will obtain in the temporally differentiated
model, we re-denote the $ \alpha $ here $ \alpha_{undiff} $.

\subsection{Case 2:  $ \lambda_1 = \lambda_2 = \lambda_3 = 1 $}

When $ \lambda_1 = \lambda_2 = \lambda_3 = 1 $, then the enzyme levels for enyzmes $ E_1 $ and $ E_2 $ are at their
maximal levels during the time interval $ [s (T_1 + T_2), s (T_1 + T_2) + T_1] $, and are not present
during the time interval $ [s (T_1 + T_2) + T_1, (s + 1) (T_1 + T_2)] $.  Therefore, in the time interval
$ [s (T_1 + T_2) + T_1, (s + 1) (T_1 + T_2)] $, we have $ n_{E_1} = n_{E_1 R_1} = n_{E_2} = n_{E_2 R_2} = 0 $.  

By periodicity and continuity it follows that, in the time interval $ [s (T_1 + T_2), s (T_1 + T_2) + T_1] $, 
we have the initial conditions $ n_{E_1} = 2 \bar{n}_1 $, $ n_{E_1 R_1} = 0 $, $ n_{E_2} = 2 \bar{n}_2 $, $ n_{E_2 R_2} = 0 $.  
Starting with these initial conditions, we have that the solution to the differential equation for $ n_{R_1, 0} $ is,
\begin{equation}
n_{R_1, 0}(\Delta t) = n_{R_1, 0}(0) e^{-k_{D, 1} \Delta t} + \frac{f_R}{k_{D, 1}} (1 - e^{-k_{D, 1} \Delta t})
\end{equation}
where $ \Delta t = t - s (T_1 + T_2) $.

The periodicity condition means that we want $ n_{R_1, 0}(T_1 + T_2) = n_{R_1, 0}(0) $, 
giving $ n_{R_1, 0}(0) = f_R/k_{D, 1} $, which then implies that $ n_{R_1, 0}(\Delta t) =
f_R/k_{D, 1} $ for all $ \Delta t $.  So for $ R_1 $, the lowest-order nonvanishing
term for $ n_{R_1} $ is unaffected by a temporal division of labor.

Turning to $ n_{E_1 R_1, 1} $, we have, in the time interval $ [s (T_1 + T_2), s (T_1 + T_2) + T_1] $,
the differential equation,
\begin{equation}
\frac{d n_{E_1 R_1, 1}}{dt} = 2 \frac{k_{11}}{V} \frac{f_R}{k_{D, 1}} - (\frac{k_{11}}{V} \frac{f_R}{k_{D, 1}} + k_{12}) n_{E_1 R_1, 1}
\end{equation}
which may be solved to give,
\begin{equation}
n_{E_1 R_1, 1}(\Delta t) = 2 \frac{A}{B} (1 - e^{-B \Delta t})
\end{equation}
where $ A $ and $ B $ were defined in the previous subsection.

The differential equation for $ n_{R_2, 1} $ takes on two distinct forms, depending on the time interval we are in. 
For $ t \in [s (T_1 + T_2), s (T_1 + T_2) + T_1], [s (T_1 + T_2) + T_1, (s + 1) (T_1 + T_2)] $, the differential
equations are, respectively,
\begin{eqnarray}
&   &
\frac{d n_{R_2, 1}}{dt} = 2 k_{12} \frac{A}{B} - 2 k_{12} \frac{A}{B} e^{-B \Delta t} - k_{D, 2} n_{R_2, 1}
\nonumber \\
&   &
\frac{d n_{R_2, 1}}{dt} = -k_{D, 2} n_{R_2, 1}
\end{eqnarray}

These equations may be solved to give,
\begin{eqnarray}
&   &
n_{R_2, 1}(\Delta t) = n_{R_2, 1}(0) e^{-k_{D, 2} \Delta t} + 2 k_{12} \frac{(A/B)}{k_{D, 2}} (1 - e^{-k_{D, 2} \Delta t})
\nonumber \\
&   &
- 2 k_{12} \frac{(A/B)}{k_{D, 2} - B} (e^{-B \Delta t} - e^{-k_{D, 2} \Delta t})
\nonumber \\
&   &
n_{R_2, 1}(\Delta t) = n_{R_2, 1}(T_1) e^{-k_{D, 2} (\Delta t - T_1)} 
\end{eqnarray}

Since we must have that $ n_{R_2, 1}(T_1 + T_2) = n_{R_2, 1}(0) $, we have,
\begin{eqnarray}
n_{R_2, 1}(0) 
& = & 
n_{R_2, 1}(T_1) e^{-k_{D, 2} T_2}
\nonumber \\
& = &
n_{R_2, 1}(0) e^{-k_{D, 2} (T_1 + T_2)} 
\nonumber \\
&   &
+ 2 k_{12} \frac{(A/B)}{k_{D, 2}} (e^{-k_{D, 2} T_2} - e^{-k_{D, 2} (T_1 + T_2)})
\nonumber \\
&   &
- 2 k_{12} \frac{(A/B)}{k_{D, 2} - B} (e^{-B T_1 - k_{D, 2} T_2} - e^{-k_{D, 2} (T_1 + T_2)})
\nonumber \\
\end{eqnarray}
and so,
\begin{eqnarray}
n_{R_2, 1}(0) 
& = & 
2 k_{12} \frac{A}{B} \frac{e^{-k_{D, 2} T_2}}{1 - e^{-k_{D, 2} (T_1 + T_2)}} 
\times 
\nonumber \\
&   &
[\frac{1 - e^{-k_{D, 2} T_1}}{k_{D, 2}}
-
 \frac{e^{-B T_1} - e^{-k_{D, 2} T_1}}{k_{D, 2} - B}] 
\end{eqnarray}

This gives,
\begin{eqnarray}
n_{R_2, 1}(\Delta t) 
& = & 
2 k_{12} \frac{(A/B)}{k_{D, 2}} 
\times \nonumber \\
&   &
[1 - (\frac{1 - e^{-k_{D, 2} T_2}}{1 - e^{-k_{D, 2} (T_1 + T_2)}}  
\nonumber \\
&   &
- \frac{k_{D, 2}}{k_{D, 2} - B} \frac{1 - e^{-(B T_1 + k_{D, 2} T_2)}}{1 - e^{-k_{D, 2} (T_1 + T_2)}}) e^{-k_{D, 2} \Delta t}
\nonumber \\
&   &
- \frac{k_{D, 2}}{k_{D, 2} - B} e^{-B \Delta t}]
\end{eqnarray}

Finally, the differential equation for $ n_{E_2 R_2, 1} $ is given by,
\begin{equation}
\frac{d n_{E_2 R_2, 1}}{dt} = 2 \frac{k_{21}}{V} n_{R_2, 1} - k_{22} n_{E_2 R_2, 1}
\end{equation}
for $ t \in [s (T_1 + T_2), s (T_1 + T_2) + T_1] $, while $ n_{E_2 R_2, 1}(\Delta t) = 0 $ for
$ t \in [s (T_1 + T_2) + T_1, (s + 1) (T_1 + T_2)] $.

Therefore, for $ t \in [s (T_1 + T_2), s (T_1 + T_2) + T_1] $, we have,
\begin{widetext}
\begin{eqnarray}
n_{E_2 R_2, 1}(\Delta t) 
& = & 
4 \frac{k_{12} k_{21}}{V} \frac{(A/B)}{k_{D, 2}} 
\times \nonumber \\
&   &
[\frac{1}{k_{22}} - \frac{1}{k_{22}} e^{-k_{22} \Delta t} 
\nonumber \\
&   &
- (\frac{1 - e^{-k_{D, 2} T_2}}{1 - e^{-k_{D, 2} (T_1 + T_2)}} 
 - \frac{k_{D, 2}}{k_{D, 2} - B} \frac{1 - e^{-(B T_1 + k_{D, 2} T_2)}}{1 - e^{-k_{D, 2} (T_1 + T_2)}})
\frac{1}{k_{22} - k_{D, 2}} e^{-k_{D, 2} \Delta t} 
\nonumber \\
&   &
+  (\frac{1 - e^{-k_{D, 2} T_2}}{1 - e^{-k_{D, 2} (T_1 + T_2)}}
- \frac{k_{D, 2}}{k_{D, 2} - B} \frac{1 - e^{-(B T_1 + k_{D, 2} T_2)}}{1 - e^{-k_{D, 2} (T_1 + T_2)}}
+ \frac{k_{D, 2} (k_{22} - k_{D, 2})}{(k_{D, 2} - B)(k_{22} - B)})
\frac{1}{k_{22} - k_{D, 2}} e^{-k_{22} \Delta t}
\nonumber \\
&   &
- \frac{k_{D, 2}}{k_{D, 2} - B} \frac{1}{k_{22} - B} e^{-B \Delta t}]
\end{eqnarray}
\end{widetext}

Note then that $ n_{E_2 R_2, 1} $ starts at $ 0 $ and then rises to a steady-state value.  Therefore,
in order to maximimize the average production rate of $ R_3 $, we must have that $ T_1 = \infty $,
so that the average production rate of $ R_3 $ is simply given by the steady-state production
rate.

When $ n_1 = n_1^{+} = 2 \bar{n}_1 $, $ n_2 = n_2^{+} = 2 \bar{n}_2 $, the steady-state value of
$ n_{E_2 R_2} $ is given by,
\begin{equation}
n_{E_2 R_2} = 4 \frac{k_{12} k_{21}}{V} \frac{(A/B)}{k_{D, 2} k_{22}} \bar{n}_1 \bar{n}_2
\end{equation}
and so, when $ T_1 $ is large, the total amount of $ R_3 $ that is produced during the time period $ T_1 $ is given
by,
\begin{equation}
\Delta n_{R_3} = k_{22} n_{E_2 R_2} T_1 = 4 \frac{k_{12} k_{21}}{V} \frac{(A/B)}{k_{D, 2}} \bar{n}_1 \bar{n}_2 T_1
\end{equation}

To ensure periodicity of the solution, $ T_2 $ must be such that the amount of $ R_3 $ consumed in the
third step is equal to the amount of $ R_3 $ accumulated.  Assuming that the amount of $ R_3 $
is infinite at all times (as with the undifferentiated case, this assumption maximizes the overall production
rate of $ P $), then the amount of $ R_3 $ consumed is given by $ 2 k_{32} \bar{n}_3 T_2 $.  We then have,
\begin{equation}
\frac{k_{12} k_{21}}{V} \frac{(A/B)}{k_{D, 2}} (2 \bar{n}_1) (2 \bar{n}_2) T_1 = k_{32} (2 \bar{n}_3) T_2
\end{equation}

Now, note that since all the enzymes are focused on either the first two tasks or the third task in the
temporally differentiated model, we have $ n = 2 \bar{n}_1 + 2 \bar{n}_2 = 2 \bar{n}_3 $.  Defining
$ \beta = 2 \bar{n}_1/n $ therefore gives,
\begin{equation}
\frac{k_{12} k_{21}}{V} \frac{(A/B)}{k_{D, 2}} \beta (1 - \beta) n^2 T_1 = k_{32} n T_2
\end{equation}

Note that the average production rate of $ P $ is simply given by $ k_{32} n T_2/(T_1 + T_2) $.
Although $ T_1 $ and $ T_2 $ are both infinite, $ T_1/(T_1 + T_2) $, $ T_2/(T_1 + T_2) $ are finite,
and we can determine the optimal split between the two work cycles that maximizes the average
production rate of $ P $.

Defining $ \alpha = T_1/(T_1 + T_2) $, we get that the average production rate of $ P $ is simply
$ k_{32} (1 - \alpha) n $.  Diving both sides of the previous equation by $ T_1 + T_2 $ gives,
\begin{equation}
\frac{k_{12} k_{21}}{V} \frac{(A/B)}{k_{D, 2}} \beta (1 - \beta) \alpha n^2 = k_{32} (1 - \alpha) n
\end{equation}

As with the steady-state solution, note that we can maximize the output of $ P $ when $ \beta = 1/2 $,
and so we obtain,
\begin{equation}
\frac{k_{12} k_{21}}{V} \frac{(A/B)}{k_{D, 2}} \alpha n = 4 k_{32} (1 - \alpha)
\end{equation}

Defining $ \gamma $ as before gives,
\begin{equation}
\alpha = \frac{1}{1 + \gamma n} \Rightarrow 1 - \alpha = \frac{\gamma n}{1 + \gamma n}
\end{equation}

To distinguish this $ \alpha $ from the $ \alpha $ defined in the temporally undifferentiated
case, we re-denote the $ \alpha $ defined in this subsection by $ \alpha_{diff} $.

\section{Comparison of System Output with and without Temporal Differentiation}

We now wish to compare the rate of production of $ P $ with and without temporal differentiation,
to determine whether temporal differentiation can optimize system performance.  In comparing
the production rate of $ P $ for both the temporally differentiated and undifferentiated cases,
it makes sense to search for long-term solutions that maximize the production of $ P $ for both
cases.  For, if we find that temporal differentiation outcompetes temporal non-differentiation,
and if the non-differentiated case is operating at a steady-state that is not optimal, then we have
not proven anything, since it is possible that a temporally non-differentiated system will produce
at least as much $ P $ as a temporally differentiated system, with the appropriate steady-state solution.
The converse holds if temporal non-differentiation outcompetes temporal differentiation, and temporal
differentiation is not running optimally.

If we expand both the undifferentiated and differentiated expressions to second-order in $ \gamma n $, 
we obtain, for small $ \gamma n $, that,
\begin{equation}
1 - \alpha_{undiff} = \gamma n (1 - 2 \gamma n)
\end{equation}
for the undifferentiated case, and,
\begin{equation}
1 - \alpha_{diff} = \gamma n (1 - \gamma n)
\end{equation}
so that temporal differentiation leads to a rate of production of $ P $ that is faster
than the undifferentiated case.  As $ \gamma n $ increases, the production advantage
for temporal differentiation increases over a certain interval.  Since both values of $ 1 - \alpha $ 
approach $ 1 $ as $ \gamma n \rightarrow \infty $, the production advantage for temporal differentiation reaches
a maximum and then disappears as $ \gamma n $ grows.

Now, plugging in the explicit definitions for $ A $ and $ B $, we obtain that,
\begin{equation}
\gamma n = \frac{k_{12} k_{21}/V}{4 k_{32} k_{D, 2}} \frac{\frac{f_R}{k_{12}}}{\frac{f_R}{k_{12}} + \frac{k_{D, 1}}{(k_{11}/V)}} n
\end{equation}

Note then that $ \gamma n $ increases with $ n $ and $ f_R $, and decreases with $ k_{D, 1} $ and $ k_{D, 2} $.

\section{Discussion}

\subsection{Justification of model parameter regimes}

Our model assumes that the total number of enzymes involved in processing the resource is small, and that the intermediate
$ R_3 $ does not decay.  The reason for the first assumption is that if the number of enzymes involved in processing the
resource is large, then temporal differentiation will not lead to an increase in the production rate of $ P $.  At large
enzyme numbers, the rate limiting step to the production rate is simply the input rate of external resource to the compartment.

We also assumed that $ k_{D, 3} = 0 $, because we claim that a low decay rate of the intermediate $ R_3 $ is a requirement
for temporal differentiation to result in a higher production rate of $ P $.  Intuitively, if the decay rate of $ R_3 $ is 
significant, then if the enzymes focus on the first two processing tasks and not on the third, the unprocessed $ R_3 $
will simply decay, so that when the enzymes switch tasks and focus on processing $ R_3 $, there will be little left.

However, when the decay rate of $ R_3 $ is low, then it makes sense for the enzymes to focus on the first two processing
tasks for a certain time period, for during that time the rate of production of $ R_3 $ is more than double what it would 
be without temporal differentiation.  Since the $ R_3 $ decays very slowly, when the enzymes switch tasks and focus on
processing $ R_3 $ into the final product $ P $, little to none of the unprocessed $ R_3 $ has decayed, so that
almost all of the $ R_3 $ produced gets converted into $ P $.  

In this paper, we assumed that $ k_{D, 3} = 0 $, which means that none of the $ R_3 $ decays.  Therefore, optimal
production rate of $ P $ is achieved with infinitely long cycle times.  When $ k_{D, 3} $ is positive (but still
small), the decay of the unprocessed $ R_3 $ means that optimal production rate of $ P $ requires a finite
cycle time.

\subsection{The small $ n $ criterion}

The analytical solution of the first four differential equations governing our model explicitly made use of the
assumption that $ \bar{n}_1 $ and $ \bar{n}_2 $ are small.  We therefore need to investigate what the criteria
for smallness are.

When $ n = \bar{n}_1 + \bar{n}_2 + \bar{n}_3 $ is small, the rate of the first reaction is given by,
\begin{equation}
k_{12} n_{E_1 R_1} = \frac{f_R}{\frac{f_R}{k_{12}} + \frac{k_{D, 1}}{(k_{11}/V)}} \bar{n}_1
\end{equation}

As $ \bar{n}_1 $ increases, eventually there will be enough enzyme $ E_1 $ present to process all of the
incoming resource.  At this point, the reaction rate becomes $ f_R $.  Therefore, the transition from
small $ \bar{n}_1 $ to large $ \bar{n}_1 $ behavior is given by the crterion \cite{DIFFTANN},
\begin{equation}
k_{12} n_{E_1 R_1} = f_R
\end{equation}
Denoting $ \bar{n}_{1, trans} $ as the value of $ \bar{n}_1 $ where the transition from small to large
$ \bar{n}_1 $ behavior occurs, we have,
\begin{equation}
\bar{n}_{1, trans} = \frac{f_R}{k_{12}} + \frac{k_{D, 1}}{(k_{11}/V)}
\end{equation}
Note that the transition point increases as $ f_R $ and $ k_{D, 1} $ increase, and as $ k_{11} $ decrease.

Now, in the small $ n $ regime, the rate of the second reaction is given by,
\begin{equation}
k_{22} n_{E_2 R_2} = \frac{(k_{21}/V)}{k_{D, 2}} \frac{f_R}{\frac{f_R}{k_{12}} + \frac{k_{D, 1}}{(k_{11}/V)}} \bar{n}_1 \bar{n}_2
\end{equation}

Setting $ \bar{n}_1 = \bar{n}_2 = \bar{n}_{12} $ for maximal reaction rate, and solving the equation $ k_{22} n_{E_2 R_2} = f_R $
for $ \bar{n}_{12} $, we obtain \cite{DIFFTANN},
\begin{equation}
\bar{n}_{12, trans} = \sqrt{\frac{k_{D, 2}}{(k_{21}/V)} (\frac{f_R}{k_{12}} + \frac{k_{D, 1}}{(k_{11}/V)})}
\end{equation}
Note that this transition point also increases as $ f_R $, $ k_{D, 1} $ and $ k_{D, 2} $ increase,
and as $ k_{11} $ and $ k_{12} $ decrease.  However, note that because of the presence of the square-root,
the dependence of $ n_{12, trans} $ on these various parameters is weaker than in the previous case.

In any event, the small $ n $ expressions that we developed in this paper are only valid when either $ f_R $,
$ k_{D, 1} $, or $ k_{D, 2} $ are large, or when $ k_{11} $ or $ k_{12} $ are small.  This general criterion
for small $ n $ makes sense:  If $ f_R $ is large compared to $ n $, then there are comparatively few agents
that can handle the incoming flow of resource.  If either $ k_{D, 1} $ or $ k_{D, 2} $ is large, then the
incoming resource or intermediate decays quickly, so that there are insufficient numbers of agents that can grab and process
the resource or intermediate before it decays.  Finally, if either $ k_{11} $ or $ k_{12} $ is small, then
the agents grab the resource or intermediate slowly, and so the resource or intermediate decays before
the agents have a chance to grab and process them.  In this case, too, there are insufficient numbers of
agents to process all of the incoming resource.

\subsection{When can a temporally differentiated process outperform a non-differentiated process?}

We now turn our attention to the problem of when a temporally differentiated pathway outperforms
a non-differentiated pathway.  

First of all, the advantage of any kind of differentiation is that, by having agents specialize
in one or a few subtasks associated with a given task, they can perform this smaller set of subtasks
much better than a non-specialized agent.  Presumably, then, this results in faster completion of the
whole task \cite{DIFFTANN}.

However, there is a cost to differentiation, due to the need to transport intermediates to the
appropriate specialized agents.  When the total number of agents is small, the transport costs are
sufficiently high that the non-differentiated pathway outperforms the differentiated pathway.  At higher
agent numbers, however, the population density of agents becomes sufficiently high that the transport
costs become sufficiently low that the differentiated pathway can outperform the non-differentiated
pathway \cite{ECONDIV5, DIFFTANN}.

The differentiated pathway can only outperform the non-differentiated pathway over a finite interval,
because the production rate of the final product becomes resource limited at sufficiently high agent
numbers.  For these agent numbers, the differentiated and non-differentiated pathways perform similarly.
Therefore, a differentiated pathway can only outperform a non-differentiated pathway when the agent-to-resource
ratio is at intermediate values.  Furthermore, the greater the flow rate of resource, the greater the likelihood
that the differentiated pathway will outperform the non-differentiated pathway.  The reason for this is that,
as the flow rate of resource grows, the agent number at which the production rate of $ P $ becomes resource limited
is pushed to higher agent numbers, and so the interval over which the differentiated pathway outperforms the
non-differentiated pathway is longer.

Thus far, our discussion has only focused on the non-temporal division of labor.  With temporal division of labor,
the agents do not specialize in only one or a few subtasks, but rather oscillate between the various subtasks of the
overall process.  In principle, given enough time, a given agent can ``learn'' the given subtask and optimize its
performance to equal that of the specialized agent.  If this optimization time is negligible, then under certain
conditions (such as those in our model), the temporally differentiated process will outperform the non-temporally
differentiated process.

If the optimization time is positive, a temporally differentiated process can outperform a
non-temporally differentiated process if the decay rate of the intermediate $ R_3 $ is $ 0 $.
The reason for this is that, for the temporally differentiated process, the optimal production
rate of $ P $ is achieved with an infinite cycle time.  Therefore, the finite optimization time
has a negligible effect on the overall production rate.

However, if $ k_{D, 3} > 0 $, then the temporally differentiated pathway will only perform optimally
with a finite cycle time.  For if the time period during which the agents focus on the first two
subtasks is infinite, then the accumulated $ R_3 $ will decay away, resulting in a low production
rate of $ P $.  In this case, the optimization time will have an effect on the overall production
rate.  Here, in order for the temporally differentiated pathway to outperform the non-temporally differentiated
pathway, the production rate of the temporally differentiated pathway with a zero optimization time must
be sufficiently greater than the production rate of the non-temporally differentiated pathway, so that the reduction 
in production rate due to a positive optimization time does not erase the superior performance of the temporally differentiated
pathway.

Now, when $ n $ is sufficiently small, we have already explained that a non-differentiated process will outperform a differentiated
one.  For larger $ n $, $ n $ may still be sufficiently small that the low-order expansions used in this paper are valid, but
large enough that the differentiated pathway can overtake the non-differentiated pathway.

If $ n $ is in either of these regimes, then at low values of $ f_R $, the value of $ \gamma n $ may be sufficiently
small that the cost associated with temporal differentiation means that it is not an optimal labor strategy (as measured
by comparing the values of $ 1 - \alpha_{undiff} $ and $ 1 - \alpha_{diff} $).

At higher values of $ f_R $, the value of $ \gamma n $ can rise to a level where the temporally differentiated
pathway with a negligible optimization time is significantly more efficient than the non-differentiated strategy.
Therefore, even with a cost arising from a positive optimization time, the temporally differentiated strategy 
outproduces the non-temporally differentiated strategy.

Thus, in the small $ n $ regime, increasing the value of $ f_R $ at a given value of $ n $ increases the 
performance advantage of the temporally differentiated labor strategy.

\subsection{Implications for sleep}

In the context of sleep, the implications of the results presented here are that sleep emerges because
the brain can process more tasks if it adopts a temporally differentiated labor strategy.  Presumably,
the more tasks a brain can accomplish within a given amount of time, the greater the survival advantage
for the organism, providing an evolutionary selection pressure for temporally differentiated labor strategies.

In our model, we have seen that an optimal ratio between the times devoted to two different sets of process
subtasks emerges, even with a constant inflow of external resource.  In the context of sleep, this suggests
a natural sleep cycle that can exist independently of any external day-night regulation.  This also suggests
an evolutionary basis for sleep that could apply to nocturnal organisms \cite{RATSLEEP}.

This being said, the presence of a day-night cycle could nevertheless regulate the exact location of the
various subtask time intervals.  Presumably, it makes sense for most organisms to remain alert during
the day, when external information is most available, and to process that information at night, when
external information is less available (the ability to avoid predators and to hunt stealthily are probably
the major selection pressures driving the emergence of nocturnal organisms) \cite{SLEEPTANK}.

As described in the Introduction, we also argue that the model presented in this paper suggests 
an evolutionary basis for the emergence of distinct REM and non-REM sleep states from an earlier
undifferentiated sleep state.  As the brain complexity increases, and the amount of information
that must be processed during the sleep state increases, it becomes more efficient for the brain
to oscillate between various information processing and consolidation subtasks associated with
the sleep state itself \cite{OLDSLEEP, SLEEPTANK}.

\section{Conclusions and Future Research}

This paper presented a highly simplified three-step model for the conversion of some external
resource into a final product.  We showed that, when the number of agents available for processing
the resource is small, and when the second intermediate does not decay, then the production rate
of the final product can be maximized if the agents oscillate between the first two subtasks
and the third subtask.  Based on these results, we conjectured that sleep and the emergence 
of REM and non-REM sleep are driven by a selective advantage for a temporally differentiated
labor strategy.

Because our model assumed that the second intermediate does not decay, we obtained an optimal
cycle time that was infinite, though the ratio of times allocated to the various sets of subtasks
during a given cycle was well-defined.  For future research, we would like to explore how the optimal
cycle time is affected when we assume a small, but positive value for $ k_{D, 3} $.  Presumably, this
will lead to an optimal cycle time that is finite, for if the cycle time is infinite, then
when the agents switch to the third task, all of the accumulated $ R_3 $ will have decayed away, so that
the production rate of $ P $ will be $ 0 $.  

In the context of biological systems, it is known that sleep oscillates between REM and non-REM states
with a well-defined cycle time.  It would be interesting to develop a model along the lines of the
model considered in this paper that, based on a few experimentally measurable parameters, could
predict the REM/non-REM cycle times via an optimization criterion.  Along these lines, it 
would be interesting to also develop a model that could similarly predict the sleep/wake cycles
of animals that have little to no exposure to the sun, such as rats (so that their sleep/wake
cycles must be internally regulated) \cite{RATSLEEP}.

We explained in the previous section that a finite cycle time, combined with an
optimization time for enzyme efficiency, can lead to distinct parameter regimes where temporal
differentiation either outperforms or underperforms non-temporal differentiation.  The characterization
of these specific parameter regimes is an issue that we plan to explore in future work as well.

Finally, in this paper we considered a fundamentally non-linear model.  This was necessary, since
a linear dependence of process rates on agent number will not give an advantage to temporal differentiation
with a mean-field description of the dynamics.  However, when agent numbers are small, 
stochastic effects can become important, since it is impossible to have fractional agents, and since it 
is often not possible to split tasks (i.e. either a task is completed or it is not).  In this situation,
even a fundamentally linear system can optimize its output via a temporally differentiated labor strategy.
The influence of stochastic effects on temporal differentiation is an issue that will also be explored
in future work.

\begin{acknowledgments}

This research was supported by the Israel Science Foundation (Alon Fellowship).

\end{acknowledgments}

\end{document}